\documentclass[10pt,a4paper]{article}
\usepackage{theorem}          %additional LaTeX2e package (see pctex\latex2e)
\usepackage{graphicx}        %graphic and .eps files

\theoremstyle{changebreak}                % (see LATEX2E\THEOREM.DTX)

\title{Performance Comparison of Function Evaluation Methods}

\author{Leo Liberti\thanks{Centre for Process Systems Engineering,
Imperial College of Science, Technology and Medicine, London,
UK. E-mail: {\tt l.liberti@ic.ac.uk}.}}

\begin{document}

\maketitle

\begin{abstract} 
We perform a comparison of the performance and efficiency of four
different function evaluation methods: black-box functions, binary
trees, $n$-ary trees and string parsing. The test consists in
evaluating 8 different functions of two variables $x,y$ over 5000
floating point values of the pair $(x,y)$. The outcome of the test
indicates that the $n$-ary tree representation of algebraic
expressions is the fastest method, closely followed by black-box
function method, then by binary trees and lastly by string
parsing. 
\end{abstract}

{\bf Keywords:}
function evaluation, tree, algebraic expression, parser

\pagestyle{myheadings}
\thispagestyle{plain}
\markboth{L. LIBERTI}{FUNCTION EVALUATION METHODS}

{\bf Important warning}. There is a mistake in the test code that
invalidates the most important result of this paper, i.e. that $n$-ary
tree based function evaluation is faster than black-box function
evaluation. It is not true: it is slower by about an order of
magnitude. However it is true that $n$-ary tree based evaluation is
faster than the other methods discussed in this paper.

\section{Introduction}
In this article we describe a test designed to measure the comparative
efficiency of four different function evaluation methods (see section
\ref{desc} for details):
\begin{itemize}
\item black-box functions;
\item binary tree representation;
\item $n$-ary tree representation;
\item string parsing.
\end{itemize}
Because of the huge number of parameters involved in such a test
(efficiency of compiler, quality of test source code, type of
hardware, type of test functions, number of variables, object code
optimization level, and so on) it is evident that this test can be neither
definitive nor undebatable. However, the results indicate a clear
winner in the $n$-ary tree representation, closely followed by
black-box functions and binary tree representation. Last (expectedly)
comes string parsing. It is somewhat of a surprise to discover that
$n$-ary tree representation gathers better results than the
precompiled black-box functions method. This finding is discussed below
(section \ref{results}).

Existing literature in this topic is scarce or non-existent. Some of
the early work focused on how to handle floating point computation
efficiently, rather than on the actual method used for the evaluation
\cite{ashenhurst}; in other instances, the evaluation of certain
classes of functions (e.g. polynomials, see \cite{fike}) was
investigated.

Most people tend to use the black-box functions method by exploiting
the compiler's capabilities in this sense. Further efforts for better
evaluation methods are usually sought after only in connection to very
specific functions and problems, e.g. in astronomy \cite{schlitt},
in number theory \cite{odlyzko}, when using discrete/boolean
functions \cite{mcgeer}.

A novel evaluation method, based on threaded binary trees, was
proposed in \cite{keeping}; this method partially eliminates the cost
of recursion by ``threading'' the binary expression tree before the
evaluation. Because the operation of threading the binary tree is
recursive in nature, the CPU time savings are only possible if the
same tree is evaluated many times (as is the case in most engineering
applications). However, the benefits of this approach should decrease
with the use of $n$-ary trees, as threading a list of like operators
in an $n$-ary tree has no effect. 

The test consists in evaluating eight different functions of one and
two variables $x,y$ over 5000 randomly generated pairs of values for
$(x,y)$, all in the interval $[0,1]$. The functions are:
\begin{enumerate}
\item $x$;
\item $x+y$;
\item $x^y$;
\item $(x+y)x^y$;
\item $\sin(x)$;
\item $\sin((x+y)x^y)$;
\item $x+y+1$;
\item $2xy(x+y+1)$.
\end{enumerate}
\label{functions}
The above functions have been chosen as a representative set for unary
and binary operators, in the sense that both binary operators (sum,
power) and unary operators (the sine function) are present. Little
does it matter for the outcome of the test that not all operator types
have been employed, for the time taken to carry out floating point
computation would have been exactly the same in all cases.

The number of variables has been limited to two for
simplicity. However, for reasons which will become apparent below (see
section \ref{narytree} about the description of the $n$-ary tree
representation), adding more variables to the expressions would only
have served to overemphasize the outcome of the tests, especially in
the case where long sequences of like operands are employed
(e.g. linear equations, or products of the kind $x_1x_2x_3\cdots
x_n$).

The test code has been written in pure C in order to minimize the
effect of compiler overhead, and compiled with the GNU C Compiler
version 2.95.2. All optimization options have been tested 
({\tt -O}, {\tt -O2}, {\tt -O3}) as well as ``debug'' mode and
``normal'' (no flags) mode. In all cases the test results have been
consistent with the order: $n$-ary trees, black-box functions, binary
trees, string parsing.

The test has been run on a Pentium-III 450MHz with 192MB RAM with the
Linux operating system. All results have been obtained by running the
executables in single-user mode and by flushing RAM caches after each
run. Enabling the caches and running in multi-user modes gathers
similar results but occasionally the black-box functions wins out (by
very little indeed) over the $n$-ary tree method. However, situations
where the black-box functions method wins over the other methods
cannot be replicated; they depend very much on the behaviour of the
caching code and on the general load of the machine at any given
moment. In any case, this kind of outcome only occurs when repeatedly
runnning the same executable over and over again as different
processes, {\it not} when calling the same function many times within
the same process.

Section \ref{desc} describes the four types of evaluation methods
tested. Section \ref{code} describes the implementation of the
methods. Section \ref{results} discusses the results of the test.
The code used to run the test can be downloaded, inspected and reused
under the GNU public license from 
\begin{quote}
 {\tt http://liberti.dhs.org/liberti/evaltest}.
\end{quote}

\section{Evaluation Methods}
\label{desc}
In this section we shall carry out a theoretical analysis of the
evaluation methods tested in this article.

\subsection{Black-box Functions}
This method for function evaluation is by far the easiest to implement
and the most commonly used within the scientific community, especially
where test code has to be rigged up or once-only jobs need to be
run. It basically lets the compiler do the work of parsing the
expression into a binary tree which is hardwired in the object code at
compile time. Evaluations are supposed to be very fast (mainly because
most of the work is carried out once only at compile time); its main
drawback is that changes to the function formulation entail
recompilation of the source code, which for most pieces of software is
not an acceptable solution.

The programming paradigm for black-box functions follows the lines of
the pseudocode below:
{\small
\begin{verbatim}
main:
  float x, y, f;
  f = blackbox(x, y);
end

function blackbox(float x, float y):
  float z;
  z = sin((x + y)*x^y);
  return z;
end function
\end{verbatim}
}

The reason why this method is called ``black-box functions'' is that
from the main procedure point of view, the function is really a black
box, in the sense that apart from knowing what argument it requires,
there is no run-time control over it.

\subsection{Binary Trees}
\begin{figure}[h]
\begin{center}
\setlength{\unitlength}{4144sp}%
\begingroup\makeatletter\ifx\SetFigFont\undefined%
\gdef\SetFigFont#1#2#3#4#5{%
  \reset@font\fontsize{#1}{#2pt}%
  \fontfamily{#3}\fontseries{#4}\fontshape{#5}%
  \selectfont}%
\fi\endgroup%
\begin{picture}(2311,2266)(308,-1509)
\thinlines
\put(1216,524){\circle{450}}
\put(541,-421){\circle{450}}
\put(1801,-421){\circle{450}}
\put(1216,-1276){\circle{450}}
\put(2386,-1276){\circle{450}}
\put(1216,299){\line(-4,-3){669.600}}
\put(1216,299){\line( 6,-5){588.688}}
\put(1801,-646){\line(-3,-2){591.923}}
\put(1801,-646){\line( 3,-2){591.923}}
\put(1171,479){\makebox(0,0)[lb]{\smash{\SetFigFont{12}{14.4}{\rmdefault}{\mddefault}{\updefault}
+
}}}
\put(496,-466){\makebox(0,0)[lb]{\smash{\SetFigFont{12}{14.4}{\rmdefault}{\mddefault}{\updefault}
$x$
}}}
\put(1756,-466){\makebox(0,0)[lb]{\smash{\SetFigFont{12}{14.4}{\rmdefault}{\mddefault}{\updefault}
$+$
}}}
\put(1126,-1321){\makebox(0,0)[lb]{\smash{\SetFigFont{12}{14.4}{\rmdefault}{\mddefault}{\updefault}
$y$
}}}
\put(2341,-1321){\makebox(0,0)[lb]{\smash{\SetFigFont{12}{14.4}{\rmdefault}{\mddefault}{\updefault}
1
}}}
\end{picture}
\end{center}
\caption{Binary tree representation for $x+y+1$.}
\label{f:binary}
\end{figure}
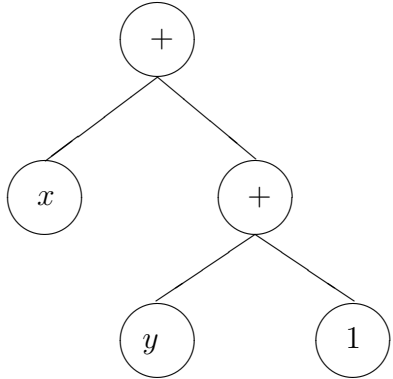

This representation is based on the idea that operators, variables and
constants are nodes of a digraph; binary operators have two outcoming
edges and unary operators have only one; leaf nodes have no outcoming
edges (for graph-related terminology and definitions, see \cite{harary},
\cite{korte}). For example, the function $x+y+1$ would be represented
as in fig. \ref{f:binary}.
Where unary operators are employed, a dummy second operand is often used.

This type of function representation is the most commonly used where
there is a need for some degree of run-time control over the
definition of the mathematical function being represented. However, it
should be noted that performing algebraic operations on this
representation is not overly simple. Most software that does not do
symbolic manipulation of algebraic expressions employs this kind of
representation. Furthermore, most general-purpose compilers (including
the GNU C compiler) use this representation too.

\subsection{$n$-ary Trees}
\label{narytree}
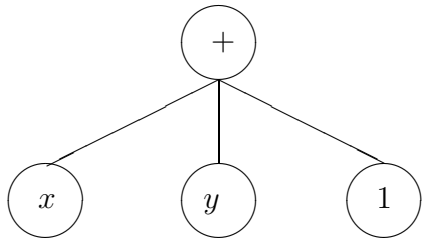
\begin{figure}[h]
\begin{center}
\setlength{\unitlength}{4144sp}%
\begingroup\makeatletter\ifx\SetFigFont\undefined%
\gdef\SetFigFont#1#2#3#4#5{%
  \reset@font\fontsize{#1}{#2pt}%
  \fontfamily{#3}\fontseries{#4}\fontshape{#5}%
  \selectfont}%
\fi\endgroup%
\begin{picture}(2491,1411)(308,-654)
\thinlines
\put(541,-421){\circle{450}}
\put(496,-466){\makebox(0,0)[lb]{\smash{\SetFigFont{12}{14.4}{\rmdefault}{\mddefault}{\updefault}
$x$
}}}
\put(2566,-421){\circle{450}}
\put(2521,-466){\makebox(0,0)[lb]{\smash{\SetFigFont{12}{14.4}{\rmdefault}{\mddefault}{\updefault}
1
}}}
\put(1576,524){\circle{450}}
\put(1531,479){\makebox(0,0)[lb]{\smash{\SetFigFont{12}{14.4}{\rmdefault}{\mddefault}{\updefault}
$+$
}}}
\put(1576,-421){\circle{450}}
\put(1486,-466){\makebox(0,0)[lb]{\smash{\SetFigFont{12}{14.4}{\rmdefault}{\mddefault}{\updefault}
$y$
}}}
\put(1576,299){\line(-2,-1){1026}}
\put(1576,299){\line( 0,-1){495}}
\put(1576,299){\line( 2,-1){990}}
\end{picture}
\end{center}
\caption{$n$-ary tree representation for $x+y+1$.}
\label{f:nary}
\end{figure}

This technique is a combination of binary trees and lists. Operators
can have any number of operands. This allows for much more efficient
handling of sequences of like operands, e.g. in linear expressions or
long products ($x_1x_2\cdots x_n$). For example, the function $x+y+1$
would be represented as in fig. \ref{f:nary}.

This type of function representation is often employed when symbolic
manipulation is required. The data structures used by languages like
Prolog and especially LISP are very similar in concept to $n$-ary
trees. 

\subsection{String Parsing}
String parsing is a process by which a string containing an algebraic
expression is evaluated directly without middle steps like tree
representation. String parsers usually consist of a lexical analyser,
which returns tokens (operators), symbols and constants, and a
grammatical interpreter which drives the lexical analyser on the given
string. It then performs the mathematical operations signalled by the
tokens on the operands. When a symbol is returned, it is looked up on
a symbol table to discover its numeric value. Because of the design
complexity, it is to be expected that string parsing is slower than
other methods. It is mainly employed where the parsing is to be
carried out once only (possibly as a pre-processing step to some main
algorithm).

By changing the grammatical interpreter, a string parser can be used
to build binary or $n$-ary trees for algebraic expressions input as
strings. This is how compilers transform source code
(i.e. strings) into object code.

\section{Implementation in the C Language}
\label{code}

The implementation of the techniques described in section \ref{desc}
above has been carried out in C in order to minimize the amount of
compiler-generated overhead code, as the C language has very low
requirements in this respect. Furthermore, no external library has
been used as it would have invalidated the timing tests. Instead,
all the code necessary to the test has been written from scratch.

The part of the test concerning black-box functions was the easiest to
code. No particular coding technique was employed. Functions returning
results of the test expressions were compiled into the executable
and called from the main routine.

Tree handling, in both the binary and $n$-ary forms, required more
work. A tree, for the purposes of this test, is defined as follows.
{\small
\begin{verbatim}
struct tree {
  int optype;          // operator type
  long varindex;       // variable index if node is a variable
  double value;        // value if node is a constant
  struct tree** nodes; // subnodes if node is not a leaf node
  long nodesize;       // number of subnodes
};
\end{verbatim}
}
The above definition is generic enough to be able to accommodate both
binary and $n$-ary trees. For binary trees, {\tt nodesize} is always
set to 2. No ``string to tree'' parser has been included as the test
code did not need that kind of generality; all the function trees
(both binary and $n$-ary) have been manually coded in. 

The string parser has been derived from the ideas given in
\cite{stroustrup}. The parser code given therein has been modified to
support exponentiation and unary functions in the form $f(x)$.

\subsection{Code Validation}
The validity of the code has been verified along with the test
proper. All results from evaluations with the four methods described
above coincide (up to at least three significant digits) for each of
the test expressions.

\section{Results}
\label{results}

As has been mentioned in the introduction, the test consists in
evaluating the eight expressions above (see page \pageref{functions}) over
5000 randomly generated pair values for $(x,y)$ (all in the interval
$[0,1]$) with each of the four described evaluation methods, trying
all the possible compiler code optimization flags. The test is carried
out in a single-tasking environment where memory cache has little or
no effect. The results of the test are reported in table \ref{t:results}.

\begin{table}
\begin{center}
\begin{tabular}{|l|r|r|r|r|r|} \hline Test Results
&\multicolumn{5}{|c|}{\bf Compiler Code Optimization Level} \\ \hline
{\bf Evaluation Method} &{\sc Normal} & {\sc Debug} ({\tt -g}) & 
{\tt -O} & {\tt -O2} & {\tt -O3} \\
\hline 
{\it Black-box functions} &0.76s & 0.76s & 0.75s & 0.74s & 0.73s \\
{\it Binary trees} &       0.84s & 0.84s & 0.80s & 0.82s & 0.82s \\
{\it $n$-ary trees} &      0.73s & 0.74s & 0.70s & 0.70s & 0.70s \\
{\it String parsing} &     2.88s & 2.97s & 2.63s & 2.23s & 2.56s \\ \hline
\end{tabular}
\end{center}
\caption{Test results. Values are expressed in seconds of ``user'' CPU
time (time spent on system calls was 0.00s in all cases).}
\label{t:results}
\end{table}

These results are surprising because normally we would expect
black-box functions to be the most efficient evaluation
method, whereas the actual ``test winner'' is the $n$-ary tree
representation (although, as has been noted in the introduction, this
test is far from definitive --- the parameters that can affect
performance are too many to be controlled all at once). This 
result is strengthened by the consideration that black-box functions
are so easy to program that the test cannot be invalidated because of
``programming errors'' or inefficient coding. On the other hand, it
may be true that with more careful coding, the results referring to
tree evaluation could be made even better.

\subsection{$n$-ary Trees: the Best Evaluation Method}
In order to explain this result, one has to consider the similarities
between black-box functions and binary trees. Although from the
programmer's point of view the two methods are far from similar, the
resulting machine code need not be all that different. As has been
explained earlier, most compilers work in such a way as to hard-code
binary trees representing the expressions within the object code. This
binary tree structure may be hidden behind a simpler logic flow than
that generated by the binary tree method, but the operations are
carried out much in the same order in both methods. Thus, it comes to
no surprise that the binary tree method gathers results which are
worse, but not by much, than those of black box functions. The two
methods are very similar, but in the black-box function case the code
is created directly by the compiler and can be better optimized.

The $n$-ary tree method performs in the same way as the binary tree
method, except where sequences of like operands with length greater
than 2 appear in the expressions (e.g. in linear expressions with at
least 3 nonzero coefficients), where it performs much faster. The
evaluation algorithm is as follows:
{\small
\begin{verbatim}
double eval(struct tree* expression, double varvalues) {
  int i;
  double ret;
  switch(expression->optype) {
  case CONSTANT:
    return expression->value;                       // leaf node
  case VARIABLE:
    return varvalues[expression->varindex];         // leaf node
  case SUM:
    ret = 0;
    for (i = 0; i < expression->nodesize; i++) {
      ret += eval(expression->nodes[i], varvalues); // recursive call
    }
    return ret;
  case PRODUCT:
    ret = 1;
    for (i = 0; i < expression->nodesize; i++) {
      ret *= eval(expression->nodes[i], varvalues); // recursive call
    }
    return ret;
  // ... all other cases
  }
}
\end{verbatim}
}
Consider the cases depicted in fig. \ref{f:binary} and \ref{f:nary}. In
the first case, where the {\tt nodesize} is always 2, a
recursive function call has to be performed for each of the two '+'
operators; in the second case, however, only one recursive function
call needs to be carried out (for the only '+' operator). 

This explains the advantage of using $n$-ary trees in evaluation of
algebraic expressions. The computational cost of generating a
recursive function call is high for most compilers. It becomes evident
that the longer ``like operand sequences'' are, the better this
evaluation method becomes.

\subsection{Results on String Parsing}
String parsing, although definitely the worst method, was also
somewhat surprising in that it was {\it not} as bad as a superficial
analysis would suggest. After all the code complexity of a lexical
analyser and a grammatical interpreter is far greater than the other
evaluation algorithms presented above. However, especially when memory
caching was allowed, the timings of the string parsing method got
better and better. The best result we obtained was close to 1.00s; so
even though it still worse than the other methods, it was the
technique that benefited most from caching (in different processes,
however, not in the same process). However, the same test carried
out using the popular {\sc Unix} utility {\tt bc} (in most
implementations based around the lex and yacc compiler tools)
gathered an appalling result of over 26s of user CPU time,
notwithstanding the fact that Stroustrup's adapted parser is highly
recursive in nature whereas lex and yacc usually generate
non-recursive (and hence theoretically faster) code.

\section{Conclusion}
In this paper we have analysed the performance of four common
evaluation methods: black-box functions, binary trees, $n$-ary trees
and string parsing. The result of the test indicates that the $n$-ary
tree expression representation is the best function evaluation method.

% end paper

\bibliographystyle{alpha}

\begin{thebibliography}{MMS{\etalchar{+}}95}

\bibitem[Ash64]{ashenhurst}
R.L. Ashenhurst.
\newblock Function evaluation in unnormalized arithmetic.
\newblock {\em Journal of the ACM}, 11(2):168--187, April 1964.

\bibitem[Fik67]{fike}
C.T. Fike.
\newblock Methods of evaluating polynomial approximations in function
  evaluation routines.
\newblock {\em Communications of the ACM}, 10(3):175--178, March 1967.

\bibitem[Har71]{harary}
Frank Harary.
\newblock {\em Graph Theory}.
\newblock Addison-Wesley, Reading, MA, 2nd edition, 1971.

\bibitem[KP97]{keeping}
B.R. Keeping and Constantinos~C. Pantelides.
\newblock Novel methods for the efficient evaluation of stored mathematical
  expressions on scalar and vector computers.
\newblock {\em AIChE Annual Meeting}, Paper \#204b, nov 1997.

\bibitem[KV00]{korte}
Bernhard Korte and Jens Vygen.
\newblock {\em Combinatorial Optimization, Theory and Algorithms}.
\newblock Springer Verlag, Berlin, 2000.

\bibitem[MMS{\etalchar{+}}95]{mcgeer}
Patrick~C. McGeer, Kenneth~L. McMillan, Alexander Saldanha, Alberto~L.
  Sangiovanni-Vincentelli, and Patrick Scaglia.
\newblock Fast discrete function evaluation using decision diagrams.
\newblock In {\em Proceedings of the Conference on Computer Aided Design},
  pages 402--407, San Jose, CA, 1995. IEEE/ACM.

\bibitem[Odl90]{odlyzko}
A.M. Odlyzko.
\newblock Primes, quantum chaos, and computers.
\newblock In {\em Number Theory}, pages 35--46. National Research Council,
  1990.

\bibitem[Sch00]{schlitt}
Wayne Schlitt.
\newblock The xstar $n$-body solver: Theory of operation.
\newblock {\em {\tt http://www.midwestcs.com/xstar/n-body}}, February 2000.

\bibitem[Str99]{stroustrup}
Bjarne Stroustrup.
\newblock {\em The C++ Programming Language}.
\newblock Addison-Wesley, Reading, MA, third edition, 1999.

\end{thebibliography}

\newcommand{\etalchar}[1]{$^{#1}$}

\end{document}